# Artificial Intelligence-Based Image Classification for Diagnosis of Skin Cancer: Challenges and Opportunities

Manu Goyal[1], Thomas Knackstedt[2], Shaofeng Yan[3], and Saeed Hassanpour[4]

*Abstract*— Recently, there has been great interest in developing Artificial Intelligence (AI) enabled computer-aided diagnostics solutions for the diagnosis of skin cancer. With the increasing incidence of skin cancers, low awareness among a growing population, and a lack of adequate clinical expertise and services, there is an immediate need for AI systems to assist clinicians in this domain. A large number of skin lesion datasets are available publicly, and researchers have developed AI solutions, particularly deep learning algorithms, to distinguish malignant skin lesions from benign lesions in different image modalities such as dermoscopic, clinical, and histopathology images. Despite the various claims of AI systems achieving higher accuracy than dermatologists in the classification of different skin lesions, these AI systems are still in the very early stages of clinical application in terms of being ready to aid clinicians in the diagnosis of skin cancers. In this review, we discuss advancements in the digital image-based AI solutions for the diagnosis of skin cancer, along with some challenges and future opportunities to improve these AI systems to support dermatologists and enhance their ability to diagnose skin cancer.

## I. INTRODUCTION

According to the Skin Cancer Foundation, the global incidence of skin cancer continues to increase [1]. In 2019, it is estimated that 192,310 cases of melanoma will be diagnosed in the United States [2]. However, the most common forms of skin cancer are non-melanocytic, such as Basal Cell Carcinoma (BCC) and Squamous Cell Carcinoma (SCC). Non-melanoma skin cancer is the most commonly occurring cancer in men and women, with over 4.3 million cases of BCC and 1 million cases of SCC diagnosed each year in the United States, although these numbers are likely to be an underestimate [3]. Early diagnosis of skin cancer is a cornerstone to improving outcomes and is correlated with 99% overall survival (OS). However, once disease progresses beyond the skin, survival is poor [4], [5].

In current medical practice, dermatologists examine patients by visual inspection with the assistance of polarized light magnification via dermoscopy. Medical diagnosis often depends on the patient's history, ethnicity, social habits and exposure to the sun. Lesions of concern are biopsied in an office setting, submitted to the laboratory, processed as permanent paraffin sections, and examined as representative glass slides by a pathologist to render a diagnosis.

AI-enabled computer-aided diagnostics (CAD) solutions are poised to revolutionize medicine and health care, especially in medical imaging. Medical imaging, including ultrasound, computed tomography (CT), and magnetic resonance imaging (MRI), is used extensively in clinical practice. In the dermatological realm, dermoscopy or, less frequently, confocal microscopy, allows for more detailed in vivo visualization of lesioned features and risk stratification [6], [7], [8], [9], [10]. In various studies, AI algorithms match or exceed clinician performance for disease detection in medical imaging [11], [12]. Recently, deep learning has provided various end-to-end solutions in the detection of abnormalities such as breast cancer, brain tumors, lung cancer, esophageal cancer, skin lesions, and foot ulcers across multiple image modalities of medical imaging [13], [14], [15], [16], [17].

Over the last decade, advances in technology have led to greater accessibility to advanced imaging techniques such as 3D whole body photoimaging/scanning, dermoscopy, high-resolution cameras, and whole-slide digital scanners that are used to collect high-quality skin cancer data from patients across the world [18], [19]. The International Skin Imaging Collaboration (ISIC) is a driving force that provides digital datasets of skin lesion images with expert annotations for automated CAD solutions for the diagnosis of melanoma and other skin cancers. A wide research interest in AI solutions for skin cancer diagnosis is facilitated by affordable and highspeed internet, computing power, and secure cloud storage to manage and share skin cancer datasets. These algorithms can be scalable to multiple devices, platforms, and operating systems, turning them into modern medical instruments [20].

The purpose of this review is to provide the reader with an update on the performance of artificial intelligence algorithms used for the diagnosis of skin cancer across various modalities of skin lesion datasets, especially in terms of the comparative studies on the performance of AI algorithms and dermatologists/ dermatopathologists. We dedicated separate sub-sections to arrange these studies according to the types of imaging modality used, including clinical photographs, dermoscopy images, and whole-slide pathology scanning. Specifically, we seek to discuss the technical challenges in this domain and opportunities to improve the current AI solutions so that they can be used as a support tool for clinicians to enhance their efficiency in diagnosing skin cancers.

[1]Department of Biomedical Data Science, Dartmouth College, Hanover, NH, USA.
[2]Department of Dermatology, Metrohealth System and School of Medicine, Case Western Reserve University, Cleveland, Ohio, USA.
[3]Section of Dermatopathology, Department of Pathology and Laboratory Medicine, Dartmouth-Hitchcock Medical Center, Geisel School of Medicine, Dartmouth College, Hanover, NH, USA.
[4]Departments of Biomedical Data Science, Computer Science, and Epidemiology, Dartmouth College, Hanover, NH, USA.

## II. ARTIFICIAL INTELLIGENCE FOR SKIN CANCER

The major advances in this field came from the work of Esteva et al. [12] who used a deep learning algorithm on a combined skin dataset of 129,450 clinical and dermoscopic images consisting of 2,032 different skin lesion diseases. They compared the performance of a deep learning method with 21 board-certified dermatologists for classification and differentiation of carcinomas versus benign seborrheic keratoses; and melanomas versus benign nevi. The performance of AI was demonstrated to be on par with dermatologists' performance for skin cancer classification. Three main types of modalities are used for the skin lesion classification and diagnosis in the work described here: clinical images, dermoscopic images, and histopathology images. In this section, we start with analysis of publicly available skin lesion datasets, and then we provide different sub-sections dedicated to the artificial intelligence solution related to each type of imaging modality.

### A. Publicly Available Datasets for Skin Cancer

1) ISIC Archive: The ISIC archive gallery consists of many clinical and dermoscopic skin lesion datasets from across the world, such as ISIC Challenges datasets [21], HAM10000 [22], and BCN20000 [23].
2) Interactive Atlas of Dermoscopy [24]: The Interactive Atlas of Dermoscopy has 1,000 clinical cases (270 melanomas, 49 seborrheic keratoses), each with at least two images: dermoscopic, and close-up clinical. It is available for research purposes and has a fee of €250.
3) Dermofit Image Library [25]: The Dermofit Image Library consists of 1,300 high-resolution images with 10 classes of skin lesions. There is a need for a licensing agreement with a one-off license fee of €75, and an academic license is available.
4) PH2 Dataset [26]: The PH2 Dataset has 200 dermoscopic images (40 melanoma and 160 nevi cases). It is freely available after signing a short online registration form.
6) MED-NODE Dataset [27]: It consists of 170 clinical images (70 melanoma and 100 nevi cases). This dataset is freely available to download for research.
7) Asan Dataset [28], [29]: It is a collection of 17,125 clinical images of 12 types of skin diseases found in Asian people. The Asan Test Dataset (1,276 images) is available to download for research.
8) Hallym Dataset [28]: This dataset consists of 125 clinical images of BCC cases.
9) SD-198 Dataset [30]: The SD-198 dataset is a clinical skin lesion dataset containing 6,584 clinical images of 198 skin diseases. This dataset was captured with digital cameras and mobile phones.
10) SD-260 Dataset [31]: This dataset is a more balanced dataset when compared to the previous SD-198 dataset since it controls the class size distribution with preservation of 10–60 images for each category. It consists of 20,600 images with 260 skin diseases.
11) Dermnet NZ [32]: Dermnet NZ has one of the largest and most diverse collections of clinical, dermoscopic and histology images of various skin diseases. These images can be used for academic research purposes. They have additional high-resolution images for purchase.
12) Derm7pt [33]: This dataset has around 2,000 dermoscopic and clinical images of skin lesions, with a 7-point check-list criteria.
13) The Cancer Genome Atlas [34]: This dataset is one of the largest collections of pathological skin lesion slides with 793 cases. It is publicly available for the research community to use.

### B. Artificial Intelligence in Dermoscopic Images

Dermoscopy is the inspection/examination of skin lesions with a dermatoscope device consisting of a high-quality magnifying lens and a (polarizable) illumination system. Dermoscopic images are captured with high-resolution digital single-lens reflex (DSLR) or smartphone camera attachments. The use of dermoscopic images for AI algorithms is becoming a very popular research field since the introduction of many large publicly available dermoscopic datasets consisting of different types of benign and cancerous skin lesions, as shown in Fig. 1. There have been multiple AI studies on lesion diagnosis using dermoscopic skin lesion datasets, which are listed below.

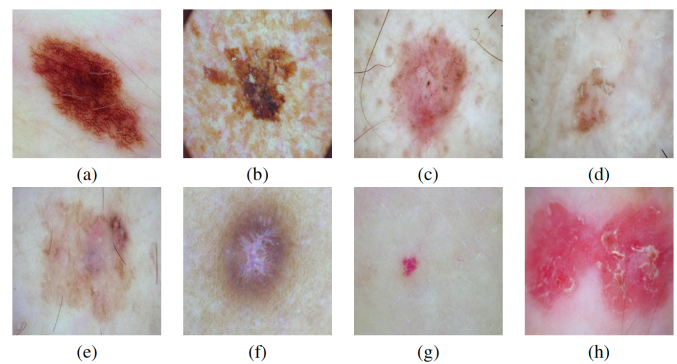

Fig. 1. Illustration of different types of dermoscopic skin lesions where (a) Nevi (b) Melanoma (c) Basal Cell Carcinoma (d) Actinic Keratosis (e) Benign Keratosis (f) Dermatofibroma (g) Vascular Lesion (h) Squamous Cell Carcinoma [22]

1) Codella et al. [35] developed an ensemble of deep learning algorithms on the ISIC-2016 dataset and compared the performance of this network with 8 dermatologists for the classification of 100 skin lesions as benign or malignant. The ensemble method outperformed the average performance of dermatologists by achieving an accuracy of 76% and specificity of 62% versus 70.5% and 59% achieved by dermatologists.
2) Haenssle et al. [36] trained a deep learning method InceptionV4 on a large dermoscopic dataset consisting of more than 100,000 benign lesions and melanoma images and compared the performance of a deep learning method with 58 dermatologists. On the test set of 100 cases (75 benign lesions and 25 melanoma cases), dermatologists had an average sensitivity of 86.6% and specificity of 71.3%, while the deep learning method achieved a sensitivity of 95% and specificity of 63.8%.

3) Brinker et al. [37] compared the performance of 157 board-certified dermatologists at 12 German university hospitals with a deep learning method (ResNet50) for 100 dermoscopic images (MClass-D) consisting of 80 nevi and 20 melanoma cases. Dermatologists achieved an overall sensitivity of 74.1%, and specificity of 60.0% on the dermoscopic dataset whereas a deep learning method achieved a specificity of 69.2% and a sensitivity of 84.2%.

4) Tschandl et al. [38] used popular deep learning architectures known as InceptionV3 and ResNet50 on a combined dataset of 7,895 dermoscopic and 5,829 close-up lesion images for diagnosis of non-pigmented skin cancers. The performance is compared with 95 dermatologists divided into three groups based on experience. The deep learning algorithms achieved accuracy on par with human experts and exceeded the human groups with beginner and intermediate raters.

5) Maron et al. [39] compared the sensitivity and specificity of a deep learning method (ResNet50) with 112 German dermatologists for multiclass classification of skin lesions which includes nevi, melanoma, benign keratosis, BCC, and SCC (also solar keratosis and intraepithelial carcinoma). The deep learning method outperformed dermatologists at a significant level ($p \leq 0.001$).

6) Haenssle et al. [40] compared the deep learning architecture based on InceptionV4 (approved as medical device by European union) and dermatologists on a dermoscopic test set consists of 100 cases (60 benign and 40 malignant lesions). This study was performed on two levels i.e. level I: dermoscopic image; level II: additional clinical close-up images, dermoscopic image, and clinical information. The deep learning algorithm achieved sensitivity and specificity score of 95% and 76.7% respectively, whereas, mean sensitivity and specificity of 89% and 80.7% respectively achieved by dermatologists in level I. With more information in level II, the mean sensitivity of dermatologists increased to 94.1% whereas mean specificity remained same.

7) Tschandl et al. [41] compared the average performance of both AI algorithms (139 in total) participated in the ISIC 2018 challenge and 511 human readers on a test set of 1511 images. In results, the AI algorithms achieved more correct diagnosis than human readers.

*C. Artificial Intelligence in Clinical Images*

Clinical images are routinely captured of different skin lesions with mobile cameras for remote examination and incorporation into patient medical records, as shown in Fig. 2. Since clinical images are captured with different cameras with variable backgrounds, illuminance and color, these images provide different insights for dermoscopic images.

1) Yang et al. [30] performed clinical skin lesion diagnosis using representation inspired by the ABCD rule on the SD-198 dataset. They compared the performance of the proposed methods with deep learning methods and dermatologists. It achieved a score of 57.62% (accuracy) in comparison to the best performing deep learning method (ResNet), which achieved 53.35%. When compared to the clinicians, only senior clinicians who have considerable experience in skin disease achieved an average accuracy of 83.29%.

2) Han et al. [28] trained a deep learning architecture (ResNet-152) to classify the clinical images of 12 skin diseases on an Asan training dataset, a MED-NODE dataset, and atlas site images, and tested it on an Asan testing set and an Edinburgh Dataset (Dermofit). The algorithm's performance was on par with the team of 16 dermatologists on 480 randomly chosen images from the Asan test dataset (260 images) and the Edinburgh dataset (220 images), whereas the AI system outperformed dermatologists in the diagnosis of BCC.

3) Fujisawa et al. [42] tested a deep learning method on 6,009 clinical images of 14 diagnoses, including both malignant and benign conditions. The deep learning algorithm achieved a diagnostic accuracy of 76.5% which is superior to the performance of 13 board-certified dermatologists (59.7%) and nine dermatology trainees (41.7%) on a 140-image dataset.

4) Brinker et al. [43] compared the performance of 145 dermatologists and a deep learning method (ResNet50) for the test case of 100 clinical skin lesion images (MClass-ND) consisting of 80 nevi cases and 20 biopsy-verified melanoma cases. The dermatologists achieved an overall sensitivity of 89.4%, a specificity of 64.4% and an AUROC of 0.769 whereas a deep learning method achieved the same sensitivity and better specificity score of 69.2%.

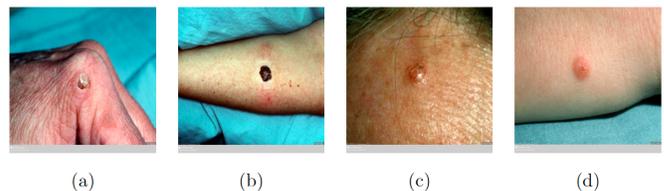

(a)　　　　　(b)　　　　　(c)　　　　　(d)

Fig. 2. Illustration of different types of clinical skin lesions where (a) Benign Keratosis (b) Melanoma (c) BCC (d) SCC [30]

*D. Artificial Intelligence in Histopathology Images*

The diagnosis of skin cancer is confirmed by dermatopathologists based on microscopic evaluation of a tissue biopsy. Deep learning solutions have been successful in the field of digital pathology with whole-slide imaging. Examples of histopathology images of skin lesions are shown in Fig. 3. These techniques are used for the classification of biopsy tissue specimens to diagnose the number of cancers such as skin, lung, and breast. In this section, we explore the deep learning methods used in digital histopathology specific to skin cancer.

1) Heckler et al. [44] used a deep learning method (ResNet50) to compare the performance of pathologists in classifying melanoma and nevi. The deep learning model was trained on a dataset of 595 histopathology images (300 melanoma and 295 nevi) and tested on 100 images (melanoma/nevi = 1:1). The total discordance with the histopathologist was 18% for melanoma, 20% for nevi, and 19% for the full set of images.

2) Jiang et al. [45] proposed the use of a deep learning algorithm on smartphone-captured digital histopathology images (MOI) for the detection of BCC. They found that the performance of the algorithm on MOI and Whole Slide Imaging (WSI) is comparable with an AUC score of 0.95. They introduced a deep segmentation network for in-depth analysis of the hard cases to further improve the performance with 0.987 (AUC), 0.97 (sensitivity), 0.94 (specificity) score.

3. Cruz-Roa et al. [46] used a deep learning architecture to discriminate between BCC and normal tissue patterns on 1,417 images from 308 Region of Interests (ROI) of skin histopathology images. They compared the deep learning method with traditional machine learning with feature descriptors, including the bag of features, canonical and Haar-based wavelet transform. The deep learning architecture proved superior over the traditional approaches by achieving 89.4% in F-Measure and 91.4% in balanced accuracy.

4) Xie et al. [47] introduced a large dataset of 2,241 histopathological images of 1,321 patients from 2008 to 2018. They used two deep learning architectures, VGG19 and ResNet50, on the 9.95 million patches generated on 2,241 histopathological images to test the classification of melanoma and nevi on different magnification scales. They achieved high accuracy in distinguishing melanoma from nevi with average F1 (0.89), Sensitivity (0.92), Specificity (0.94) and AUC (0.98).

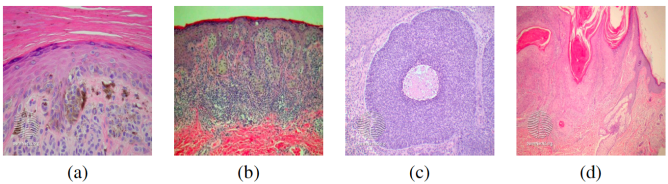

Fig. 3. Illustration of different types of histopathology images where (a) Nevi (b) Melanoma (c) Basal Cell Carcinoma (d) Squamous Cell Carcinoma [19]

## III. CHALLENGES IN ARTIFICIAL INTELLIGENCE

With deep learning algorithms surpassing the benchmarks of popular computer vision datasets in a short period, the same trend could be expected in the skin lesion diagnosis challenge as well. However, as we further explore the skin lesion diagnosis challenge, this task appears to be not straightforward like ImageNet, PASCAL-VOC, MS-COCO challenges in a non-medical domain [48], [49]. There are intra-class similarities and inter-class dissimilarities regarding color, texture, size, place, and appearance in the visual appearance of skin lesions. Deep learning algorithms generally require a substantial amount of diverse, balanced, and high-quality training data that represent each class of skin lesions to improve diagnostic accuracy. For skin lesion datasets of various modalities, there are many more issues related to the diagnosis of skin cancer with AI solutions as discussed below.

### A. Performance of Deep Learning and Unbalanced Datasets

The performance of deep learning algorithms mostly depends on the quality of image datasets rather than tuning the hyper-parameters of networks, as is commonly seen in the different publicly available skin lesion datasets. There are generally more cases of benign skin lesions rather than malignant lesions. Most of the deep learning architectures are designed on a balanced dataset, such as ImageNet, which consists of 1,000 images per class (1000 classes) [48]. Hence, the performance of a deep learning algorithm usually suffers from unbalanced datasets, despite using tuning tricks like a penalty for false negatives found in minor skin lesion classes during training using custom loss functions.

### B. Curious Case of Histopathology Images/ Digital Pathology

The size of images in clinical and dermoscopic skin lesion datasets varies between 1200 × 768 and 3648 × 2736 depending on the camera used. Most of the deep learning algorithms are usually developed and validated on large datasets of non-medical background. These deep learning algorithms have worked very well on clinical and dermoscopic skin lesion datasets by fine tuning the algorithms through transfer learning techniques. On the other hand, histopathological scans consist of millions of pixels and their dimensions are commonly larger than 50,000 x 50,000. Hence, there are many technical challenges for deep learning or AI algorithms in digital pathology such as lack of labeled data, infinite pattern from different types of tissues, high-quality feature extraction, high computational expenses and many more [50].

### C. Patients' Medical History and Clinical Meta-data

Patients' medical history, social habits, and clinical meta-data are considered when making a skin cancer diagnosis. It is very important to know the diagnostic meta-data, such as patient and family history of skin cancer, age, ethnicity, sex, general anatomic site, size and structure of the skin lesion, while performing a visual inspection of a suspected skin lesion with dermoscopy. Hence, only image-based deep learning algorithms used for the diagnosis of skin cancer falter on key aspects of patient and clinical information. It is proven in a previous study [36] that both 'beginners' and 'skilled' dermatologists' performance is improved with the availability of clinical information and that they performed better than deep learning algorithms. Unfortunately, both patient history and clinical meta-data are missing in the most publicly available skin lesion datasets.

### D. ABCDE Rule and Time-line Datasets

In the clinical setting, a suspicious lesion is visually inspected with the help of dermoscopy. The ABCDE rule is considered an important rule for differentiating benign moles (nevi) from melanoma. This includes whether the lesion is asymmetrical, has irregular borders, displays multiple colors, whether the diameter of the lesion is greater than six millimetres, and if there has been any evolution or change in the composition of the lesion. Despite the availability of dermoscopic datasets of skin lesions, deep learning algorithms do not work the same way or look for a pattern similar to the ABCDE rule trusted by clinicians. It is mainly due to the complexity of pattern recognition for the characteristics of skin cancers in medical imaging. That is why, despite recent attempts by researchers to demystify the working of deep learning algorithms, such efforts are still considered as a black-box approach, especially in medical imaging. Since there are no timeline dermoscopic datasets available publicly, it is not possible to determine the change of a lesion's characteristics according to the evolution of the ABCDE rule.

*E. Biopsy is a Must*

Although, in various studies, AI solutions outperformed the human experts in diagnosis of skin cancer. But, even if skin cancer is confirmed by AI solutions with a high confidence rate, a biopsy and histological test must still be undertaken to confirm a diagnosis, similar to real clinical practice by dermatologists and general practitioners. The diagnostic accuracy of deep learning algorithms could be misleading as well. For example, if a testing set consists of 20 melanoma and 80 nevi cases, and the overall diagnostic accuracy is 90% (100% in nevi and 50% in melanoma cases), it is dangerous for a deep learning algorithm to be used in this case as a means to deliver a diagnosis of melanoma. As misdiagnosis of a cancer patient by a deep learning algorithm could risk a fatality, a biopsy should be taken to ensure safety and confirm the algorithm's diagnosis.

*F. Inter-class Similarities (Mimics of Skin Lesions)*

A number of skin lesions can mimic skin cancer in both clinical and microscopic settings, which could result in misdiagnosis. For example, in clinical and dermoscopic images, seborrheic keratosis can mimic skin cancers including BCC, SCC, and melanoma. In histopathology images, there are many histologic mimics of BCC such as SCC, benign follicular tumors, basaloid follicular hamartoma, a tumor of follicular infundibulum, syringoma, and microcystic adnexal carcinoma [51]. Hence, deep learning algorithms, when trained on limited classes of skin lesions in a dataset, do not reliably distinguish skin cancers from their known mimics.

*G. Intra-class Dissimilarities*

Several skin lesions have intra-class dissimilarities in terms of color, attribute, texture, size, site. Hence, these skin lesions are further categorized into many sub-categories based on visual appearance. For example, the color of most melanomas is black because of the dark pigment of melanin. But certain melanomas are found to be of normal skin color, reddish, and pinkish looking. Similarly, BCC has many subcategories, such as nodular BCC, superficial BCC, morphoeic BCC, Basosquamous BCC, and their appearance is completely different from each other, ranging from white to red in color as shown in Fig. 4.

*H. Communication Barrier between AI and Dermatologists*

Sometimes even experts in computer vision find it hard to understand the decisions made by deep learning frameworks. For example, if there is an algorithm that is 85% accurate for the diagnosis of skin cancer, it is often very difficult to understand why the algorithm is making wrong inferences on the rest of the 15% of cases, and how to improve those decisions. These algorithms are not usually similar or representative of the ways in which clinicians make such decisions. Hence, deep learning algorithms are often deemed as a black box solution that does not offer clear explanation for its conclusion, and often they only provide an output in confidence probability ranging from 0 to 1 for classification of each skin lesion in a test set. Currently, it is not clear how dermatologists would interpret deep learning models' outcomes in diagnosing of skin cancer

*I. Noisy Real-life Data with Heterogeneous Data Sources*

In the current datasets of skin lesions, the dermoscopic images are captured with high-resolution DSLR cameras and in an optimal environment of lighting and distance of capture. A deep learning algorithm trained on these high-quality dermoscopic datasets, achieving a reasonable diagnostic accuracy, could potentially be scaled to smart-phone vision applications. When this model is tested on multiple smart phone captured images by different cameras in different lighting conditions and distances, the same diagnostic accuracy is hard to achieve. Deep learning algorithms are found to be highly sensitive to which camera devices are used to capture the data, and their performance degrades if a different type of camera device is used for testing. Patient-provided self-captured skin images are frequently of low-quality and are not suitable for digital dermatology [52], [53].

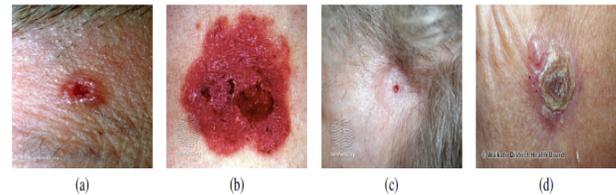

Fig. 4. Illustration of intra-class dissimilarities in BCC (a) Nodular BCC (b) Superficial BCC (c) Morphoeic BCC (d) Basosquamous BCC [32]

*J. Race, Ethnicity and Population*

Most of the cases in the current skin lesion datasets belong to fair-skinned individuals rather than brown or dark-skinned persons. Although the risk of developing skin cancer is indeed relatively high among the fair-skinned person population, people with dark skin can also develop skin cancer and are frequently diagnosed at later stages [54]. Skin cancer represents 4 to 5%, 2 to 4 %, and 1 to 2% of all cancers in Hispanics, Asians, and Blacks, respectively [55]. Hence, deep learning frameworks validated for the diagnosis of skin cancer in fair-skinned people has a greater risk of misdiagnosing those with darker skin [56]. In a recent study, Han et al. [28] trained a deep learning algorithm on an Asan training dataset consisting of skin lesions from Asians. They reported an accuracy of 81% on the Asian testing set, whereas they reported an accuracy of only 56% on the Dermofit dataset, which consists of skin lesions of Caucasian people. Therefore, this drop-in accuracy signifies a lack of transferability of the learned features of deep learning algorithms across datasets that contain persons of a different race, ethnicity, or population.

*K. Rare Skin Cancer and Other Skin Conditions*

BCC, SCC and melanoma collectively comprise 98% of all skin cancers. However, there are other skin cancers, including Merkel cell carcinoma (MCC), appendageal carcinomas, cutaneous lymphoma, sarcoma, kaposi sarcoma, and cutaneous secondaries, that are ignored by most algorithms. Beside these rare skin cancers, there are certain other skin conditions, such as ulcers, skin infections, neoplasms, and non-infectious granulomas, that could mimic skin lesions. If deep learning algorithms are trained on datasets that do not have adequate cases of these rare skin cancers and other mentioned skin conditions, there is a high risk of misdiagnosis when it is tested on these skin conditions.

*L. Incomplete Diagnosis Pipeline for Artificial Intelligence*

In the clinical setting, the diagnosis of skin cancer is made by inspecting the skin lesion with or without dermoscopy, followed by confirmatory biopsy and pathological examination. The major issue with the current publicly available skin lesion datasets is that they lack complete labels related to the diagnosis performed by a dermatologist. Nevertheless, most of the classification labels for dermoscopic skin lesion images are determined by pathological examination. Still, these dermoscopic and clinical skin lesion datasets do not have corresponding pathological classification labels to develop a complete diagnosis pipeline for AI.

## IV. OPPORTUNITIES

AI researchers invariably claim their systems exceed the performance of dermatologists for the diagnosis of skin cancer. But this picture is far from reality, as these experiments are performed in closed systems with a defined set of rules. With the many challenges mentioned in the above section, the nature of these reported performance evaluations is nowhere near the real-life diagnosis performed by clinicians treating skin cancer. Often, deep learning algorithms are deemed as opaque, as they only learn from pixel values of imaging datasets and do not have any domain knowledge or perform logical inferences to establish the relationship between different types of skin lesions [56]. But, in the future, deep learning could do very well for the diagnosis of skin cancer with the given opportunities listed below.

*A. Balanced Dataset and Selection of Cases*

A balanced dataset is critical for the good performance of deep learning algorithms used for classification tasks. Hence, balanced datasets are required with a selection of cases that completely represent the category of that particular skin lesion, and the input of experienced dermatologists could be very helpful for this selection.

*B. Computer Aided Diagnosis for Digital Pathology*

The evolution of whole slide imaging for digital image analysis and GPU clusters for computational power in recent years has attracted the interests of both pathologist and computer vision societies for developing computer-aided diagnosis in digital pathology. One of the most popular AI approach to tackle the dimensionality obstacle is to use the deep learning algorithms as sliding window classification and aggregates those classifications to infer predominant histologic patterns [15].

*C. Color Constancy for Illumination and Heterogenous Data Sources*

In publicly available clinical and dermoscopic datasets, the skin lesion images are acquired with different illumination settings and acquisition devices which could reduce the performance of the AI systems. It is proven in many studies that color constancy algorithms such as Shades of Gray, max-RGB can be used to improve the performance of AI algorithms for the classification of multisource images [57], [58]. The example of Shadow of Gray algorithm as a pre-processing method to normalize the illumination and lighting effect on dermoscopic skin lesions images, as shown in Fig. 5.

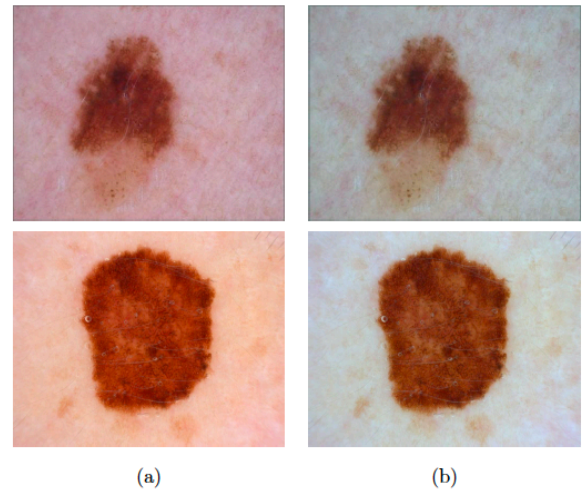

Fig. 5. Examples of pre-processing with a Shades of Gray algorithm. (a) Original images with different background colors; and (b) Pre-processed images with more consistent background colors.

*D. Diverse Datasets*

Deep learning networks are often criticised for social biases due to most of the imaging data belonging to fair-skinned persons. Skin lesion datasets need to have racial diversity, i.e., they must add equally distributed skin lesion cases from fair-skinned and dark-skinned people to reduce social or ethnic bias in deep learning models. The same concern can be extended to age, especially when the degree of skin aging or surrounding solar damage can influence the dataset and decision-making.

*E. Data Augmentation*

Data augmentation techniques may mitigate many limitations of datasets, such as unbalanced data among the classes of skin lesions and heterogeneous sources of data, by adding augmented samples with different image transformations, such as rotation, random crop, horizontal and vertical flip, translation, shear, color jitter, and colorspace. It is proven in many studies that data augmentation improved the diagnosis of skin cancer [59], [60]. In the HAM10000 dataset, [22] the skin lesion images were captured at different magnifications or angles or with different cameras, a process known as natural data augmentation. Notably, Goyal et al. [61] used a deep learning architecture called Faster R-CNN to develop an algorithm to generate augmented copies similar to the natural data augmentation method used for other skin lesion datasets.

*F. Generative Adversarial Networks*

Generative Adversarial Networks (GAN) are deep learning architectures that are attracting interest in the medical imaging community. GAN is mainly used to generate high-quality fake imaging data to overcome a limited dataset [62], [63], [64]. For skin cancer, GAN can be used to generate realistic synthetic skin lesion images to overcome the lack of annotated data [65]. The distribution of skin lesions in publicly available datasets is heavily skewed by each class's prevalence among patients, and GAN can be used to generate imaging data for under-represented skin lesion classes or rare classes of skin cancer, such as MCC, sebaceous carcinoma, or kaposi sarcoma.

*G. Identifying Sub-categories*

There could be many visual intra-class dissimilarities in the appearance of skin lesions in terms of texture, color, and size. In most of the publicly available datasets, the collection of skin lesions belongs to each superclass rather than dividing them into sub-categories. Dealing with many intra-class dissimilarities and inter-class similarities (mimics of a skin lesion) in the skin lesion dataset, it is challenging for deep learning algorithms to classify or differentiate such lesions. As a possible solution to deal with this issue, sub-categories of each skin lesion should be treated as different classes in the dataset used for training deep learning algorithms. However, this will require a greater volume of training images and it will also be more challenging to translate into clinical practice. Therefore, subcategorization would require a certain degree of suspicion or a reasonable pre-test probability to adequately aid the clinician choosing the algorithm.

*H. Semantic Explanation of Prediction*

To assist clinicians, deep learning algorithms need to provide a semantic explanation rather than just a confidence score for the prediction of skin lesions. One possible solution could be for deep learning networks using longitudinal datasets to provide a semantic explanation of networks' prediction according to ABCDE criteria (asymmetry, border, color, diameter, evolution) or 7-point skin lesion malignancy checklist (pigment network, regression structures, pigmentation, vascular structures, streaks, dots and globules, blue whitish veil) [33].

*I. Multiple Models for Diagnosis of Skin Cancer*

Rather than relying on a single AI solution for the diagnosis of skin cancer, multiple deep learning models can evaluate different features or aspects of skin lesions, submit predictions, and generate a final conclusion. In this regard, cloud computational power and storage is becoming more affordable and it will be possible to host multiple models to assist dermatologists in the diagnosis of skin cancer, around the world in parallel (or in synchrony).

*J. Combining Clinical Information and Imaging Features*

Clinical meta-data and patient history are considered clinically important in the diagnosis of skin cancer. This information can provide insight beyond the imaging features used by deep learning algorithms. Hence, there is a need to develop data fusion algorithms that can combine features comprised of clinical information with imaging features from deep learning models to provide final predictions of the diagnosis of skin cancer. In a recent study, Pacheco et al. [66] combined deep learning models (clinical images) and patient clinical information to achieve approximately 7% improvement in the balanced prediction accuracy.

*K. Multi-modality solution: Complete Diagnosis Pipeline*

If corresponding histopathological data for dermoscopic skin lesions were available in datasets, we could develop a complete AI solution similar to a dermatologist's diagnosis pipeline. In the first step, an AI solution is used to classify dermoscopic skin lesions, with a deep learning algorithm trained on the dermoscopic dataset. For suspected cases, the deep learning algorithm can be developed on a pathological dataset to determine whether the lesion is cancerous or not.

*L. Rigorous Clinical Validation*

It is a well-known fact, for both clinicians and AI researchers, that mistakes can inform future decision-making. Since we cannot afford misdiagnosis by technology, it is better to keep AI solutions in the background for rigorous validation of noisy data coming from real patients and for improving the predictions of these technological systems to date, until they are finally validated to provide useful insights into the diagnosis of skin cancer and assist clinicians either in hospital and remote settings.

V. CONCLUSION

Research involving AI is making encouraging progress in the diagnosis of skin cancer. Despite the various claims of deep learning algorithms surpassing clinicians' performance in the diagnosis of skin cancer, there are far more challenges faced by these algorithms to become a complete diagnostic system. Because such experiments are performed in controlled settings, algorithms are never tested in the real-life diagnosis of skin cancer patients. The real-world diagnosis process requires taking into account a patient's ethnicity, skin, hair and eye color, occupation, illness, medicines, existing sun damage, the number of nevi, and lifestyle habits (such as sun exposure, smoking, and alcohol intake), clinical history, the respond to previous treatments, and other information from the patient's medical records. However, current deep learning models predominantly rely on only patients' imaging data. Moreover, such systems often risk a misdiagnosis whenever they are applied to skin lesions or conditions that are not present in the training dataset. This paper further explores opportunities to build robust algorithms to assist clinicians in the diagnosis of skin cancer. Computer vision and dermatologist societies need to work together to improve current AI solutions and enhance the diagnostic accuracy of methods used for the diagnosis of skin cancer. AI has the potential to deliver a paradigm shift in the diagnosis of skin cancer, and thus a cost-effective, remotely accessible, and accurate healthcare solution.

**Search strategy and selection criteria**

We used Google Scholar and PubMed to find relevant manuscripts. We restricted our search to papers published in English between Jan 1, 2012, and Jan 11, 2020. We used the following terms in different combinations: "skin cancer", "skin lesions", "rare skin cancer", "deep learning", "artificial intelligence", "dermatologists", "clinical images", "dermoscopic", "histopathology", "social bias", "artificial intelligence and skin cancer", "skin cancer and deep learning", "dermatologists and deep learning", "skin cancer datasets", "lesion diagnosis and deep learning", "clinical information, deep learning and skin cancer", "data augmentation and skin cancer", "GAN and skin cancer".

**Author Contributions**
All authors have read and approved the manuscript, and each author has participated sufficiently in developing the project and the manuscript. M.G. contributed to the literature review and analysis of the study and drafting the manuscript. T.K., and S.Y. contributed to clinical aspects of technical challenges and opportunities and drafting the manuscript. S.H. oversaw the technical details and analysis of the study. T.K. and S.H. reviewed the manuscript.

**Funding**
This research was supported in part by National Institute of Health grants R01LM012837 and R01CA249758.

**Conflict of Interest**
Declarations of interest: none.